\documentclass[aps,prl,reprint,groupedaddress,showpacs,nofootinbib]{revtex4-1}
\usepackage{graphicx}
\usepackage{dcolumn}
\usepackage{bm}

\input{epsf}

\begin{document}
\title{Microwave absorption in a 2D topological insulators with a developed network of edge states.}

\author{M.M. Mahmoodian}
\email{mahmood@isp.nsc.ru}
\affiliation{Rzhanov Institute of Semiconductor Physics, Siberian Branch, Russian Academy of Sciences, Novosibirsk, 630090 Russia}
\affiliation{Novosibirsk State University, Novosibirsk, 630090 Russia}

\author{M.V. Entin}
\email{entin@isp.nsc.ru}
\affiliation{Rzhanov Institute of Semiconductor Physics, Siberian Branch, Russian Academy of Sciences, Novosibirsk, 630090 Russia}
\affiliation{Novosibirsk State University, Novosibirsk, 630090 Russia}

\date{\today}

\begin{abstract}
The 2D HgTe quantum well is analyzed based on the assumption that the width fluctuations convert the system to a random mixture of domains with positive and negative energy gaps. The borders between ordinary and topological insulator phases form a network of the edge states covering the overall sample. The optical transitions within the edge states yield the 2D absorption. The qualitative consideration is based on the model of optical intraedge transitions in curved edge states together with the percolation arguments.
\end{abstract}


\maketitle
\section{Introduction.}

2D topological insulators (TI) based on HgTe layers have recently attracted much attention due to their unique properties: insulating insides with conducting edges \cite{kvon,bhz,vp,qi,shen}. In this approach, the electron transport occurs near the TI borders. This transport should be one-dimensional and non-local \cite{kvon} if one does not include the backscattering processes which are strongly forbidden by the spin conservation. However, the experimental observations show that many aspects of this picture contradict the ideal picture. In particular, this concerns the absence of the 2D transport and the backscattering on the edge states.

The edge states of a topological insulator have linear spectrum $\varepsilon_k=\pm \hbar vk$, where $v$ is the edge electron velocity, $k$ is wavevector. They are formed on the boundary between the phases with negative and positive energy gap signs. Usually, the edge states are considered in the so-called Bernevig-Hughes-Zhang (BHZ) model \cite{bhz} as the states on the external border of the TI. The Volkov-Pankratov model (VP) \cite{vp}, oppositely, is based on the 2-band approximation. It gives the interface states forming near the line where the gap changes its sign. In any case, the edge states exponentially decay from the edge with some decrement $\lambda$ (main unidirectional decrement in BHZ model and bidirectional decrement in VP model).

In real 2D TI the randomness exists, that essentially changes the properties of the system. Different aspects of random TI were studied earlier. For example, Anderson topological insulator have been studied in \cite{li,groth,girschik}, based on the BHZ or Dirac models with a random on-site energy. These systems conserve the topological character in some domains of parameters with conductance quantization. The paper \cite{jiang} considered this system as a TI with the on-site randomness and it supports the conservation of its topological features.

The other study \cite{hsu} deals with the conductance fluctuations in a disordered TI and interplay between a metal, quantum spin-Hall insulator, and an ordinary insulator. All these approaches are unified by the presence of the diagonal disorder. In particular, in the envelope-function approximation, the randomness is caused by impurities. Such disorder can only alters the states of an ideal TI and  produce new states if the disorder is strong enough. In \cite{mag-ent} the interplay between the one-dimensional localization and delocalization in the edge states of a narrow quantum wire of TI, due to weak impurity backscattering, was considered.

In contrast with the cited papers, we will deal here with the situation when the chaos affects the energy gap sign. In fact, such consideration is actual, because the gap value in a 2D HgTe quantum well strongly depends on the quantum well width. Moreover of that, in experiment \cite{kvon}, width $w$ is close to critical value $w_0=6.3$ nm, when the energy gap goes through zero. Important, is that the topologically protected VP edge states appear near the lines where the gap changes its sign. The width fluctuations lead to multiple internal edges in a sample \cite{ent-mah-mag,mah-mag-ent,curve}, while the external border of TI in these systems plays a less essential role.

We will concentrate on the microwave absorption in a random 2D topological insulator. As a result of width fluctuations, the HgTe quantum well should consist of alternating domains of an ordinary insulator (OI) and TI. If the mean well width $\overline{w}$ is much less than $w_0$, the system is OI with a rare inclusion of the TI phase. If $\overline{w}\gg w_0$, rare OI domains are incorporated into TI. If $\overline{w}\to w_0$, OI and TI domains are mixed at the approximately equal proportion (see Fig.~1).

The internal borders between the phases produce a random network of the edge states. Near the threshold $|\xi-\xi_c|\ll1$, where $\xi_c=0.5$, the mixture of OI and TI phases becomes developed, and the edges between the phases cover the entire sample. The 2D conductivity of the system is conditioned by the transport along this network. We shall consider the edge states as linear objects neglecting their width.

Note that absorption coefficient $\alpha$ should be proportional to the conductivity $\sigma$ of the sample via relation $\alpha =4\pi\sigma/c$, where $c$ is the light velocity. In this relation the conductivity can depend on the frequency.

The Fermi level is assumed to lie within the overall gap. We shall use the known assumption that phototransitions occur inside branches $\varepsilon_k=\hbar vk$ or $\varepsilon_k=-\hbar vk$ only and that is valid in the VP model.

If the fluctuations have large in-plane sizes, the system can be treated as a classical random network of one-dimensional conductors. Below the localization threshold the system contains TI or OI quantum dots embedded into the opposite phase. In the threshold vicinity, the network of conductors covers all the sample and the low-temperature conductivity is established.

Different approaches are used: namely, static conductivity along the random edges, static conductivity accounting backscattering due to the tunneling between edges, active-reactive transport on low frequencies, absorption due to the edge curvature with the accounting for the edge electron quantization.

The edge electron quantization changes the picture of absorption. In this case, the phototransitions obtain the minimal critical frequency. This frequency becomes very small when the edge becomes long enough in the percolation threshold vicinity.

The edge channels geometry  is important for the optical absorption at zero temperature. We analyse this geometry. Then we shall obtain the 2D intraedge optical absorption in the assumption that they occur due to the similar mechanism as in the straight edge. However, the edge curvature leads to a possibility of intra-branch transitions due to the nonhomogeneous acting electric field. In addition, the specific conductivity mechanism caused by the active-reactive transport is discussed. The paper is completed by the Conclusion section.

\begin{figure}[ht]
\centerline{\epsfysize=1.1cm\epsfbox{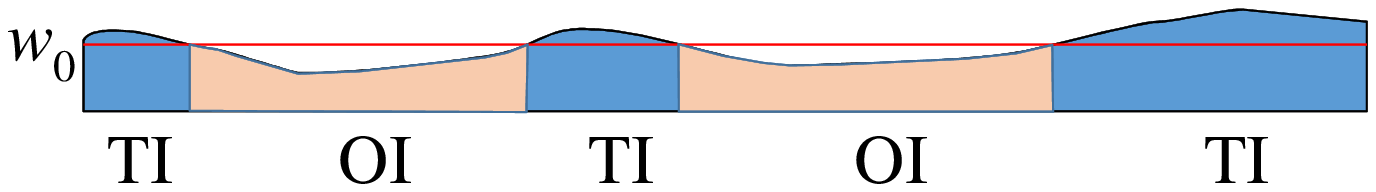}}\vspace{0.1cm}
\centerline{\epsfysize=2.1cm\epsfbox{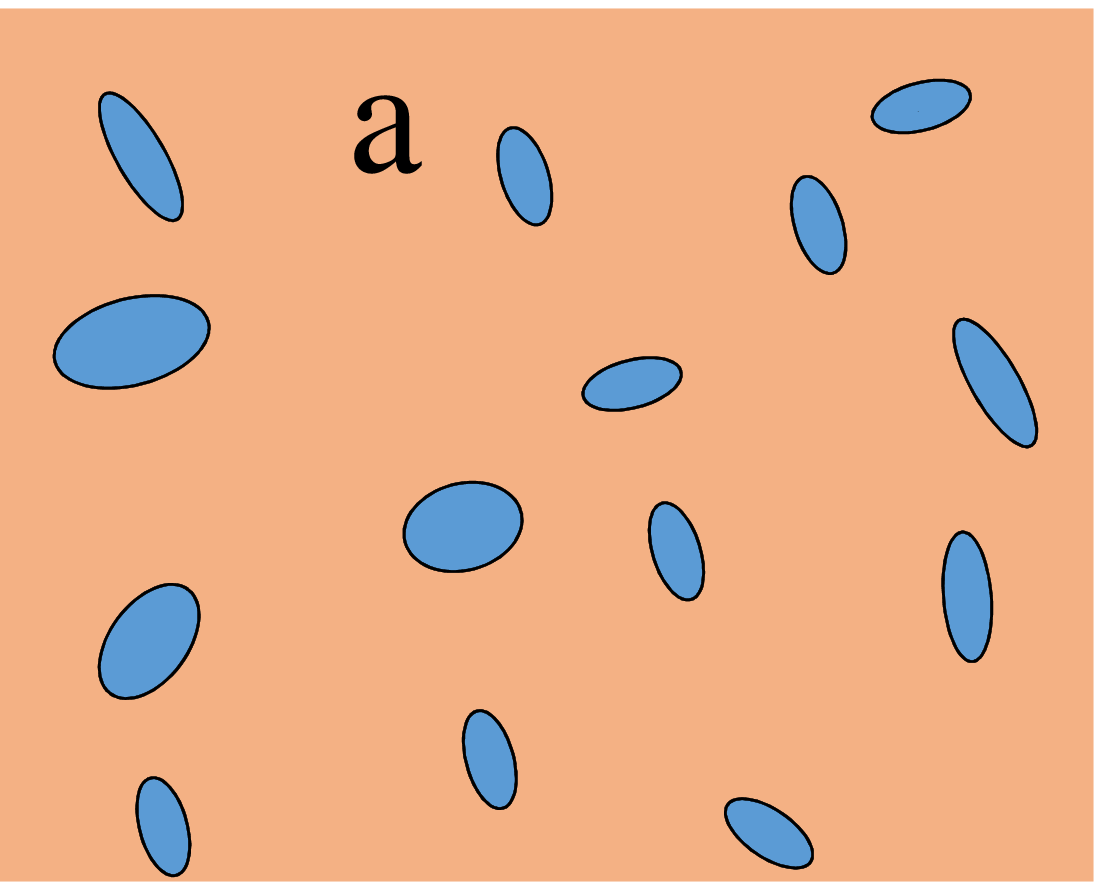}\hspace{0.1cm}\epsfysize=2.1cm\epsfbox{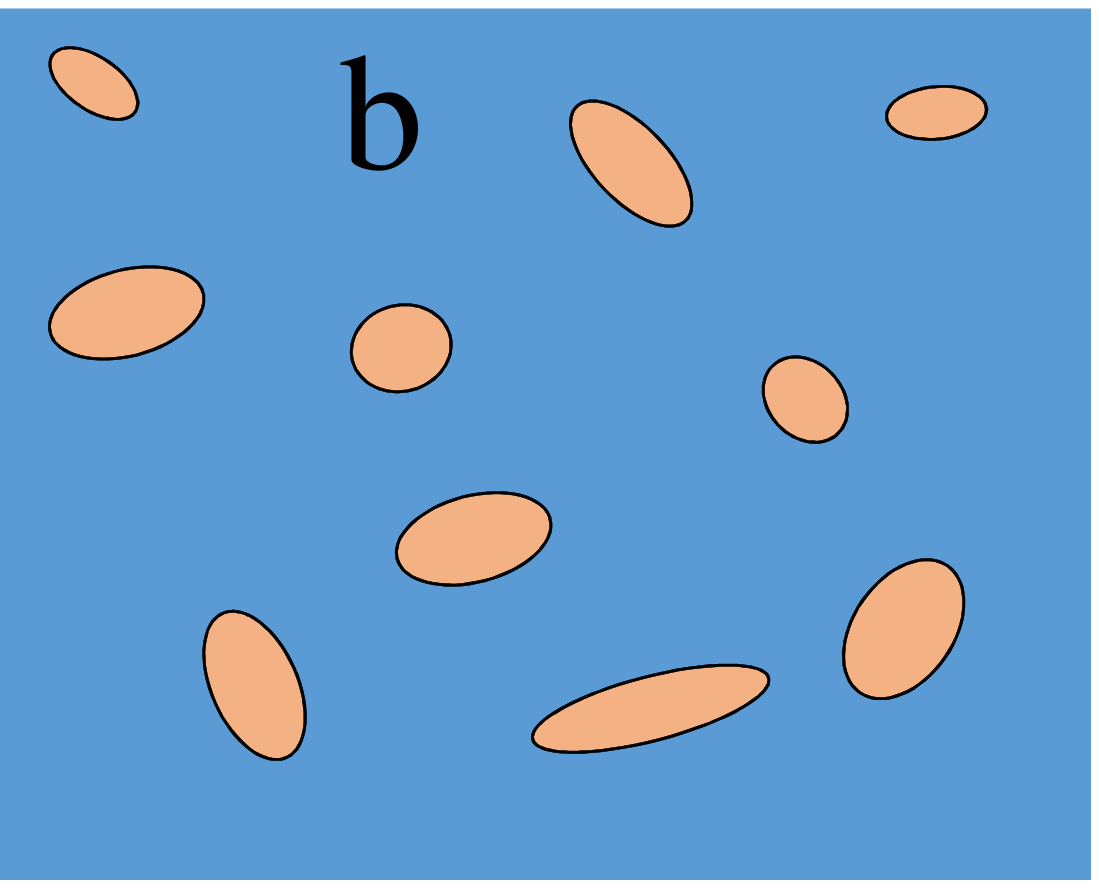}\hspace{0.1cm}\epsfysize=2.1cm\epsfbox{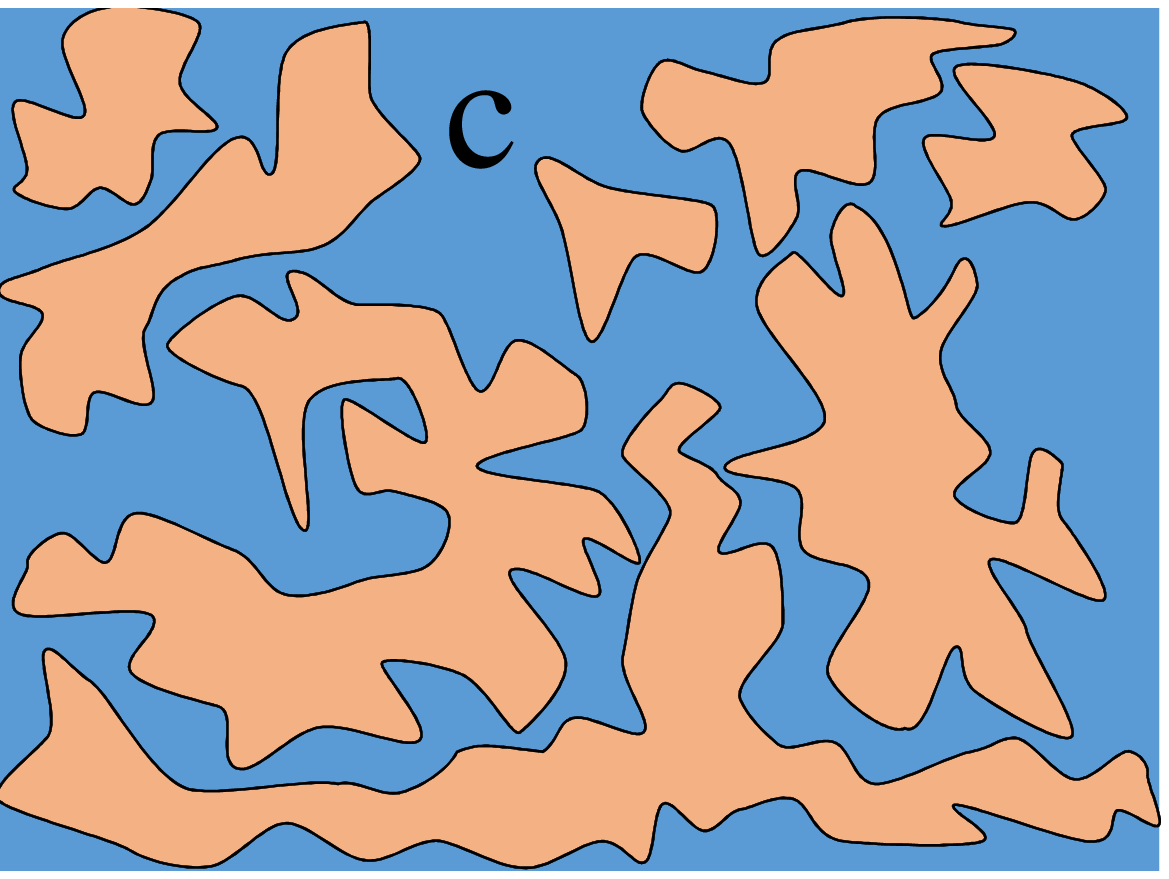}}
\caption{Rough HgTe film. The domains with widths $w<w_0$ are OI, with widths $w>w_0$ - the TI. Fig.~1a. $\xi\ll 1$: Rare TI domains (blue) embedded into the OI matrix (brown). Fig.~1b: $1-\xi\ll 1$: Rare OI domains embedded into the TI matrix. Fig.~1c: $|\xi-\xi_c|\ll1$. A developed network of OI and TI domains. The edges cross the entire sample.}\label{fig1}
\end{figure}

\section{Geometry of the edge states.}

Let the well width have a distribution function $\Pi(w)$ around the mean value $\overline{w}$,
\begin{equation}\label{fpi}
\Pi(w)=\frac{1}{\delta w}F\left(\frac{w-\overline{w}}{\delta w}\right),~~~\int\Pi(w) dw=1,
\end{equation}
where $F(x)$ is a dimensionless function.

We assume that the width fluctuations are weak, $\delta w\ll \overline{w}$, so that the integration in Eq.~(\ref{fpi}) may be continued from $-\infty$ to $\infty$. We shall characterize the system by a dimensionless parameter $\xi=\int_{w_0}^\infty\Pi(w) dw$, $0<\xi<1$. Then, if $\xi\to 0$, the OI phase prevails, if $\xi\to 1$, the TI prevails. For example, as a model, we can assume the Gaussian fluctuations, when $F(x)\propto \exp(-x^2/2)$. Then
\begin{eqnarray}\label{i}
~~~\xi\propto \exp(-(\overline{w}-w_0)^2/2\delta w^2),~~~~~~~~~\mbox{if}~~ \overline{w}<w_0,\nonumber\\
~~~1-\xi\propto \exp(-(\overline{w}-w_0)^2/2\delta w^2),~~~\mbox{if}~~ \overline{w}>w_0,
\end{eqnarray}
and $|\overline{w}-w_0|\gg\delta w$.

The edge states network of a random-gap-sign system covers the whole sample. Hence, the edge contribution to any kinetic or optical property converts to the 2D property. Naturally, this intensive property will be determined by the edge length per unit sample area.

The overall edges length is determined by the length of the $\Delta(x,y)=0$ line; in other words, by an overall perimeter of TI clusters $P(\xi)$.

In a sample with a size $L$ exceeding critical length, $P(\xi)\propto L^2$ \cite{voss}, in other words, $P(\xi)=(L^2/a) p(\xi)$, where $a$ is a characteristic spacial size of fluctuations, $p(\xi)$ is a dimensionless function not depending on the system size. The quantity $a$ in our problem should be chosen as a characteristic mean-spacial gap distribution size; for example, it can be chosen as $a=\delta w/\sqrt{\langle|\nabla w|^2\rangle}$. Evidently, the perimeter $p(\xi)\propto\xi$ at $\xi\to 0$, $p(\xi_c)\propto 1$ and $p(\xi)\propto 1-\xi$ at $\xi\to 1$. These relations can be combined  as $p(\xi)\sim\xi(1-\xi)$ at any $\xi$ (but apart from the threshold $\xi_c$).

When the system approaches percolation threshold $\xi_c$, $|\xi-\xi_c|\ll 1$, the critical cluster size infinitely grows. Below we shall use the language of the discrete site model of percolation in the square lattice \cite{shkl-efros}. Then $a$ can be treated as a lattice period. Denote the number of sites in a finite cluster by $s$. Correlation length $L_c$ behaves as $L_c=a|\xi-\xi_c|^{-\nu}$, where $\nu\approx1.34$. The geometry of critical clusters inside $L_c$ (in a square $L\times L$, at $L\ll L_c$) becomes complicated. The cluster edge length is proportional to cluster area $a^2s$, where $s$ is a number of sites occupied by the cluster. Since, near $\xi_c$, the number of all occupied sites is half of the total number of sites; in other words, in a square $L\times L$, the overall edge length $\propto L^2$. The edge lines on the scale less than correlation length $L_c$, but larger than the minimal size $a$, are fractal lines which length $P\propto L^{D_h}\propto s^{D_h/2}$ has a fractal dimension $D_h\approx 1.74$ \cite{voss}.

Another important quantity is a width of the edge state $1/\lambda$. We will assume that $1/\lambda\ll a$ so that the system becomes a geometrical network of 1D conductors. The adiabatic picture of the electron motion along edges fails if two edges are closer than $1/\lambda$ and an electron jumps from one edge to the other.

\section{Light absorption due to intra-edge-state transitions.}

Let us consider the absorption due to the transitions between the edge states network. Apparently, absorption coefficient $\alpha$ should be proportional to $p(\xi)$ and the absorption coefficient of the straight edge per unit length $\alpha_1$,
\begin{equation}\label{alfa2}
\alpha=\alpha_1 p(\xi).
\end{equation}
Value $\alpha_1$ can be extracted from \cite{mag-ent1}. However, for the straight edges, the transitions within the edge states are forbidden, unless the unusual processes including virtual transitions with the participation of 2D states \cite{mag-ent1} are taken into account. At the same time, the transitions between the edge states and the 2D states have their low-frequency threshold \cite{mah-mag-ent}. Hence, Eq.~(\ref{alfa2}) is valid until the frequency is low enough, but not too low due to the necessity for including the quantization of electron motion along the edge.

Let us take into account the edge curvature. Unlike the straight edge, in a curved edge, another intrabranch absorption mechanism exists. Namely, the acting component of light electric field is its projection onto the local edge tangent ${\bf {\bm\tau}}(u)$, where $u$ is the curvilinear coordinate along the edge. The nonuniformity of the resulting acting field makes the intrabranch transitions permitted. A simple calculation can be done with the use of the golden rule. The Hamiltonian of the interaction with homogeneous electric field ${\bf E}e^{i\omega t}=-\dot{\bf A}(t)/c$ is $H_{int}=-(e/c)v{\bf A {\bm\tau}}(u)$. The 1D wave functions of the edge states are $e^{iku}/\sqrt{as_p}$, where $s_p$ denotes the number of edge sites in a discrete model. They have the energy $\hbar vk$, where wavevector $k$ is quantized by condition $kas_p=2\pi(m+1/2)$, $m$ is integer \cite{curve}. The transition probability between these states reads
\begin{equation}\label{fw}
W_{kk'}=\frac{2\pi}{\hbar^2}|\langle k'|H_{int}|k\rangle|^2\delta(v(k-k')-\omega)).
\end{equation}

Apart from the percolation threshold, when $\xi(1-\xi)\ll 1$, the inclusions are small and, as a rule, of elliptical shape. Without the generality limitations in a suitably turned coordinate system, the edge can be expressed as ${\bf r}=R(\cos t,\tilde{e}\sin t)$ ($\tilde{e}$ is eccentricity, $R\sim a$) and after an affine transform to a circle ${\bf r}=R(\cos t,\sin t)$. It can be found that the light absorption for intraedge transitions occurs between the neighboring quantized edge states and the frequency satisfies relation $\omega=v/R$. If $\omega\gg v/a$, the absorption is proportional to $p(\xi)\propto \exp(-(\overline{w}-w_0)^2/2\delta w^2)$. The absorption is forbidden for frequencies less than $v/a$.\footnote{Note that low-frequency absorption is permitted due to extremely (exponentially) low-probable fluctuations with a large in-plane size.}

In the case of circle shape inclusions the 1D wave functions of the edge states are $e^{i m\varphi}/\sqrt{2\pi}$. The transition probability between these states for linearly polarized light reads
\begin{eqnarray}\label{fwc}
~~~~~~~~W_{mm'}=\frac{\pi}{2}\left(\frac{eEv}{\hbar\omega}\right)^2\left(\delta_{m,m'+1}+\delta_{m,m'-1}\right)\times\nonumber\\
\delta\left(\frac{v}{R}(m-m')-\omega\right).
\end{eqnarray}
The ac conductivity can be found using the relation
\begin{equation}\label{fse0}
\sigma E^2=\frac{p(\xi)}{a^2}\sum\limits_{mm'}\hbar\omega W_{mm'}f_{m}(1-f_{m'}).
\end{equation}

Let us introduce the distribution function $\Omega(R/a)/a$ for edge radii $R$, where $\int\Omega(R/a)dR/a=1$. Function $\Omega(x)$ decays at least as $1/x^\varrho, ~~\varrho>1$ for convergency.\footnote{In fact, such situation is typical if $\xi(1-\xi)\ll1$, when $\Omega(x)$ becomes exponentially small for a large $x$.} After averaging Eq.~(\ref{fse0}) over radius $R$ at low temperature we obtain
\begin{equation}\label{fs0}
\alpha=4\pi^2 p(\xi)\alpha_0F\left(\frac{v}{a\omega}\right),~~~\alpha_0=\frac{e^2}{\hbar c},~~~ F(x)=x^3\Omega(x).
\end{equation}
At $x\sim 1$, $F(x)\sim 1$. Due to improbable fluctuations with a large edge length (exponentially unlikely), the absorbtion coefficient tends to zero for a small $\omega$.

In fact, the low-frequency absorption becomes much stronger when $\xi$ approaches the percolation threshold. In a TI or OI cluster with perimeter $as_p$ the distance between levels is $2\pi\hbar v/as_p$. Hence, to obtain the absorption on frequency $\omega$, one should have a sufficiently large quantity $s_p$. The density of such clusters $n_p(s_p)$ is small.

On the analogy with the power-law dependence of the clusters with $s$ sites, $n_s$, quantity $n_p(s_p)$ has also its power-like behavior near the percolation threshold $n_p(s_p)\propto s_p^{-r}$ and the exponential decay far from the threshold. Quantity $r$ should be larger than 2 to make the total number of edge sites $\sum_{s_p>1}^\infty n_p(s_p)$ converging. The total absorption is the absorption of edges with length $as_p$, proportional to this length and the absorption of an edge per unit length $\alpha_0$. To obtain the total absorption coefficient, we should sum partial absorption coefficients with the above-mentioned limitation:
\begin{equation}\label{rfr}
\alpha(\omega)=\sum_{v/a\omega}^\infty\alpha_0s_p\frac{1}{s_p^r}=\frac{\alpha_0}{r-2}\left(\frac{a\omega}{2\pi v}\right)^{r-2}.
\end{equation}
Eq.~(\ref{rfr}) yields a power-like drop of the absorption coefficient with the frequency. The upper limit in Eq.~(\ref{rfr}), in fact, is finite for finite $|\xi-\xi_c|$. Introducing the power limitation $s_p<|\xi-\xi_c|^{-z}$ we have limit $a\omega>2\pi v|\xi-\xi_c|^z$ for the validity of Eq.~(\ref{rfr}). For $a\omega<2\pi v|\xi-\xi_c|^z$, $\alpha(\omega)$, obviously, exponentially tends to zero.

\section{Low-temperature d.c. conductivity.}

In the percolation threshold vicinity the edge states form a developed network of 1D bonds covering the overall sample. At a low temperature, each 1D bond has its d.c. conductance of the order of minimal metallic conductivity $e^2/h$. It is determined by the collisionless motion along a long edge. This gives the finite conductance of a large $L\times L$ sample of the order of conductance quantum $e^2/h$ and infinite conductivity $\sigma\propto L e^2/h$. However, the exponentially weak backscattering \cite{mag-ent}, together with a very large edge length leads to the finite conductivity, when the edge length exceeds backscattering length $l_{back}$. For blocks with size $l_b$ the edge length is $l_b^2\lambda$. Equating this quantity to $l_{back}$ we find $l_b=\sqrt{l_{back}/\lambda}$.

The conductivity estimation can be done assuming that the system can be devided into blocks $l_b\times l_b$ the conductance of which has an order of $e^2/h$. They are connected by the backscattering processes to the entire 2D network by interconnections with the same conductance. This obviously yields the overall conductivity of the same order, and $\alpha(\omega)\sim \alpha_0$. This result is valid exactly on the threshold. Apart from the threshold the limitation includes parameter $\xi$.

This reasoning is valid while the edge states quantization is not taken into account. The estimation is also applied to find the 2D microwave absorption.

\section{Reactive-active transport.}

Another factor limiting the conductivity becomes the capacity between the parts of the edge which are far from each other along the edge, but close in real space, when the interedge transport results from the interedge capacity. In principle, this factor should result in the frequency-dependent conductivity.

Consider the system near the percolation threshold. Let us separate the sample by blocks with size $l_1\times l_1$, where $l_1\ll L_c$ in such a way that the block impedance is greater than the conductance. The edges cover all the block and have area $l_1\times l_1$. Part of them is connected to the right side of the block, the other one - to the left one. At the same time they approach each other by the minimal distance $a$. The capacity of block $C$ can be estimated as determined by its area $l_1^2$: $C=4\pi l_1^2\epsilon/a$. The edge has internal conductance $e^2/h$ sequently connecting the other edge by the capacitance. The capacitive current $\propto i\omega C$ of this block should be comparable to the ohmic current $\propto e^2/h$ that determines size $l_1$, where the reactive and active currents are comparable if $l_1$ remains to be much less than $L_c$:
\begin{equation}\label{fo}
\omega C=e^2/h,
\end{equation}
\begin{equation}\label{fl}
l_1=\sqrt{e^2a/(4\pi h\epsilon\omega)}.
\end{equation}
As a result,
\begin{equation}\label{fsi}
\sigma_{l_1}=e^2/h\times \sqrt{i}.
\end{equation}
Note that the real and the imaginary parts of conductivity have the same order and do not depend on the frequency. The same estimations can also be made in the two-sort two-dimensional model with the equal density of Dykhne \cite{dykhne}.

At first glance, Eq.~(\ref{fo}) has a strange character: the conductivity including capacitive transport does not depend on frequency at low frequencies. In fact, we have considered the situation just at the percolation threshold when the size of clusters with collisionless transport (and infinite conductivity) unlimitedly grows. The limitation is brought by the deviation from the threshold: when the blocks size becomes larger than the correlation length, this mechanism fails. That determines the lowest frequency limit of the applicability of Eq.~(\ref{fo}): $\omega\gg e^2|\xi-\xi_c|^{2\nu}/(4\pi h\epsilon a)$.

If the frequency tends to 0, $l_1$ grows and exceeds $l_b$. As a result, blocks $L_c\times L_c$ become connected capacitively only. Their intercapacity can be estimated as $C_i=4\pi\epsilon l_1/a$. The resulting impedance is determined by the addition of $1/\sigma_{l_1}$ and $1/i\omega C_i$. Finally, the reactive conductivity becomes proportional to the frequency.

\section{Conclusions.}

The basic idea that integrated the consideration is the fact that 2D topological insulators have their width close to the OI-TI transition threshold. Inevitable width fluctuations convert the system to the random  mixture of TI-OI phases in which the edge states appear.

We are interested in the 2D absorption at a low frequency and temperature. The edge states cover the entire sample and the entire energy gap; this gives birth to the 2D conductivity along the edge state network.

The finite frequency gives a possibility of the 2D absorption due to distributed edges. The considered mechanisms include the interbranch microwave transitions in locally straight edges, the absorption due to active-reactive a.c. conductivity and the intrabranch absorption caused by the edges curvature. All these mechanisms result in power-like frequency dependence of the absorption at a low frequency.

It should be emphasized that the presence of the network determines the possibility of metallic state of the system when the Mott-Anderson criterium of delocalization is fullfilled. The consideration of this question and the problem of hopping transport at a finite temperature goes beyond the frames of the present paper and will be considered elsewhere.

\paragraph*{\bf Acknowledgments.} This research has been supported in part by RFBR grants No 17-02-00837.

\end{document}